\documentclass[runningheads]{llncs}
\usepackage{graphicx}
\usepackage{cite}
\usepackage{amsmath,amssymb,amsfonts}
\usepackage{algorithmic}
\usepackage{graphicx}
\usepackage{xcolor}
\usepackage{soul}
\usepackage{diagbox}
\usepackage[normalem]{ulem}
\useunder{\uline}{\ul}{}
\usepackage{mathtools}
\DeclarePairedDelimiter{\abs}{\lvert}{\rvert}
\usepackage{url}
\usepackage{paralist}
\usepackage{upgreek}
\usepackage{subfigure}
\newcommand{\etal}{~\textit{et~al.}}

\begin{document}
\title{SD-LayerNet: Semi-supervised retinal layer segmentation in OCT using disentangled representation with anatomical priors}

\author{Botond Fazekas \and
Guilherme Aresta \and
Dmitrii Lachinov \and
Sophie Riedl \and
Julia Mai \and 
Ursula Schmidt-Erfurth \and
Hrvoje Bogunovi\'c}

\institute{Christian Doppler Laboratory for Artificial Intelligence in Retina, Department of Ophthalmology and Optometry, Medical University of Vienna, Austria\\\email{\{botond.fazekas,hrvoje.bogunovic\}@meduniwien.ac.at}}

\authorrunning{B. Fazekas et al.}

\titlerunning{SD-LayerNet: Semi-supervised retinal layer segmentation in OCT}

\maketitle              
\begin{abstract}
Optical coherence tomography (OCT) is a non-invasive 3D modality widely used in ophthalmology for imaging the retina. Achieving automated, anatomically coherent retinal layer segmentation on OCT is important for the detection and monitoring of different retinal diseases, like Age-related Macular Disease (AMD) or Diabetic Retinopathy. However, the majority of state-of-the-art layer segmentation methods are based on purely supervised deep-learning, requiring a large amount of pixel-level annotated data that is expensive and hard to obtain. With this in mind, we introduce a semi-supervised paradigm into the retinal layer segmentation task that makes use of the information present in large-scale unlabeled datasets as well as anatomical priors. In particular, a novel fully differentiable approach is used for converting surface position regression into a pixel-wise structured segmentation, allowing to use both 1D surface and 2D layer representations in a coupled fashion to train the model. In particular, these 2D segmentations are used as anatomical factors that, together with learned style factors, compose disentangled representations used for reconstructing the input image. In parallel, we propose a set of anatomical priors to improve network training when a limited amount of labeled data is available. We demonstrate on the real-world dataset of scans with intermediate and wet-AMD that our method outperforms state-of-the-art when using our full training set, but more importantly largely exceeds state-of-the-art when it is trained with a fraction of the labeled data.

\end{abstract}
\section{Introduction}

Optical Coherence Tomography (OCT) is currently the main imaging modality for monitoring retinal disease and guiding treatment. With it, ophthalmologists can assess several imaging biomarkers including retinal layer thickness, an important maker for pathologies like Age-related Macular Degeneration (AMD), a leading cause of blindness worldwide\cite{2004_Bressler}. 
However, the clinicians' workload and the large size of the acquired volumes make manual layer delineation almost prohibitive in clinical practice. Therefore, automated retinal layer segmentation methods have been pursued as soon as OCTs became widely available.

Initial automated layer segmentation approaches, such as the IOWA Reference Algorithms \cite{2006_Li, 2014_Zhang}, relied on hand-crafted features, graph algorithms, and dynamic programming to estimate surface positions. These methods can fulfill hard constraints such as topological ordering, layer smoothness, or prior layer thickness. However, the reliance on hand-crafted image features hinders generalization to severe or rare pathologies, image acquisition noise, or artifacts.
Over the past years, the emergence of deep learning-based (DL) segmentation models significantly reduced the need for handcrafted features. 
One of the first DL approaches \cite{2017_Roy} used a U-Net \cite{2015_Ronneberger_CONF} to classify each pixel of the input image as one of 9 retinal layers, background or possibly fluid-filled pockets, which may lead to segmentation of layers on anatomically implausible locations. This can be circumvented with an edge detection network that predicts only a single location for a layer in an A-scan\cite{2021_Sousa}. However, none of these two approaches accounted for hard anatomical constraints. 

Contrary to this, the method of He\etal \cite{2019_He_CONF, 2021_He} outputs per layer a pixel-wise labeling and a softmax mapping that encodes the most probable location of the layer boundary, with the two representations being decoupled. Layer ordering is imposed by predicting the positions of the shallower surface and iteratively rectifying all subsequent surfaces to ensure that they have higher depth, which significantly improved the state-of-the-art.

Despite the near expert-level performance of segmentation methods on healthy OCT scans or scans with common pathologies, current approaches struggle with severe or rare pathologies for which they often require a large number of annotated samples for training. However, acquiring annotated data is time-consuming and costly. Semi-supervised approaches address this problem by exploiting the abundantly available unlabeled data. A limited number of proposed methods for retinal layer segmentation make use of semi-supervised training by using a second network to generate an adversarial loss~\cite{2019_Liu,2018_Sedai_CONF, 2019_Sedai_CONF}, but fail to incorporate retinal anatomical priors. 
A promising solution is to use disentangled learning, which has already been successfully applied to other medical imaging tasks.
In particular, Chartsias\etal \cite{2019_Chartsias} proposed a disentangled representation learning model for cardiac image analysis, called Spatial Decomposition Network (SD-Net). The network creates a strictly disjoint representation of binary spatial anatomical information and non-spatial modality-dependent style. Subsequently, these two factors are used to reconstruct the original image. With a few labeled instances and an optional adversarial loss function, a subset of the anatomical factors can be encouraged to represent meaningful anatomical regions.

In this paper, we propose Spatial Decomposition Layer Segmentation Network (SD-LayerNet), a fully convolutional semi-supervised retinal layer segmentation approach that allows to significantly reduce the amount of annotated training data required to achieve state-of-the-art performance by using disentangled representations and anatomical priors. Specifically, our main contributions are
\begin{inparaenum}
    \item a \textbf{fully differentiable topological engine}, that allows converting surface positions to pixel-wise structured segmentations, while guaranteeing the correct topological ordering, enabling the coupled use of both 1D and 2D representations of the retinal layers to train the network;
    \item a set of tailored retinal \textbf{anatomical priors} encoded as self-supervised loss terms, which improve the performance when training with very limited amounts of labeled data; and
    \item evaluation of the method on a large clinical dataset with challenging \textbf{real-world intermediate and wet-AMD} cases.
\end{inparaenum}

\section{Model architecture}

\begin{figure}[tb]
\centering
\includegraphics[width=0.99\textwidth,page=1]{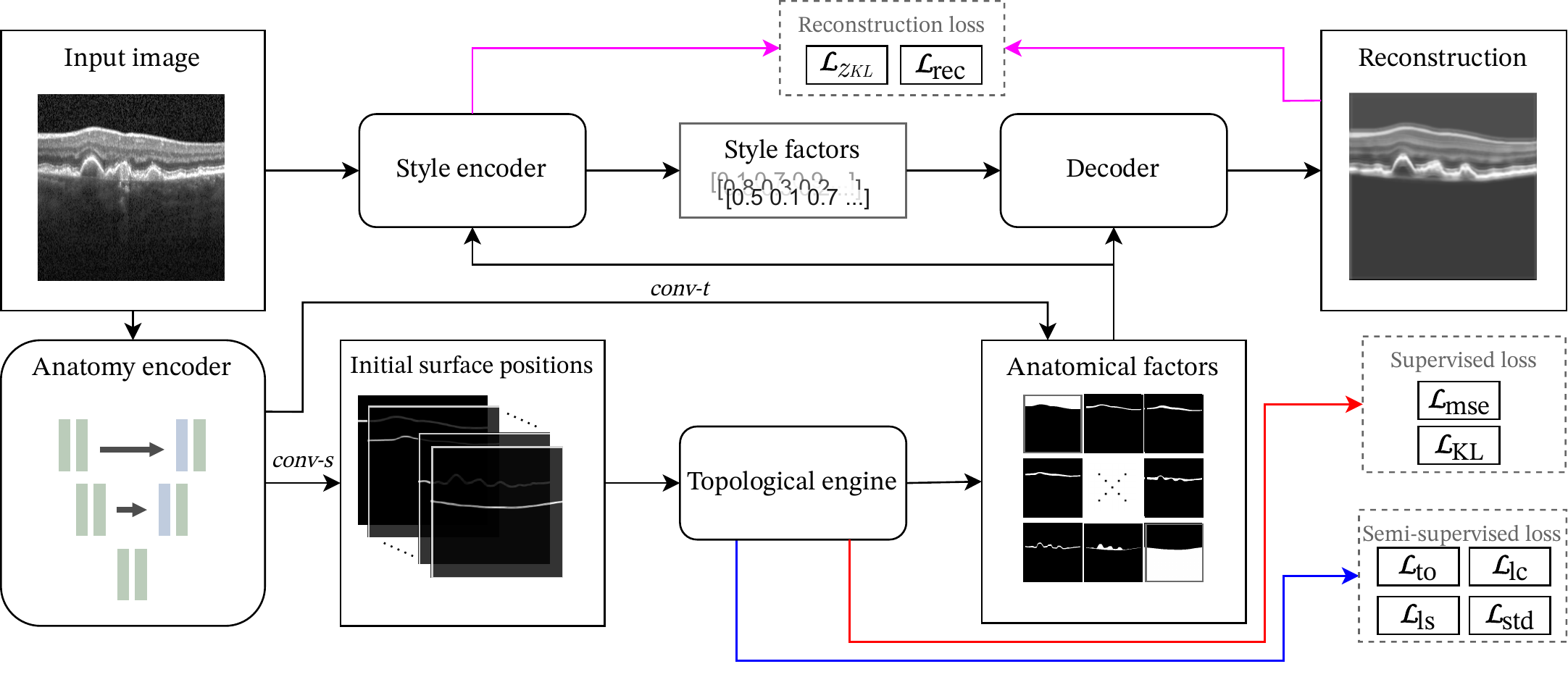}
\caption{Our network has two input branches. First, an anatomy encoder with two output branches is used for feature extraction. One output is directly used as a single spatial texture factor, while the other generates a probability map of the layer positions. A topological engine guarantees the correct ordering of the layers and generates the spatial maps of the layer. The second input branch of the network is connected to a style encoder that generates the style factors, which encodes the intensity values of the anatomical factors. The decoder generates the reconstructed image from the style and anatomical factors. Depending on whether layer annotations are available for the input image, a supervised loss (red arrow) or a self-supervised loss (blue arrow) is minimized. Regardless of the availability, the reconstruction loss (purple arrow) is always applied.} 
\label{fig:graphical-abstract}
\end{figure}

Our model (Fig.~\ref{fig:graphical-abstract}), uses both annotated and non-annotated images to learn to segment $S$ retinal surfaces on OCT image by reconstructing input image from disentangled representation \cite{2019_Chartsias}. In particular, an anatomy encoder infers the initial surface positions and one textural factor. The topological engine converts the surface positions into 2D layer representations, to form with the textural factors the anatomical factors. These spatial factors, together with inferred style factors that encode gray-scale intensity, compose the disentangled representations from which the input image is reconstructed.

\subsection{Anatomical factor generation}
A Residual Attention U-Net \cite{2020_Chen} with four encoder and four decoder stages and PReLU activation is used as the feature extractor for the anatomical factor generation. The network has two output branches: 
\begin{inparaenum}
\item \emph{conv-s} for surface position regression and
\item \emph{conv-t}, for texture factor generation.
\end{inparaenum}
The texture factor branch \emph{conv-t} has a single sigmoid activated channel output, which is concatenated to the anatomical factors created from the surface position regressions, and is aimed at capturing the speckle noise present in the scans.

\subsubsection{Layer position regression}

Let $W$ and $H$ be the width and height of the input OCT slice (B-scan) with S surfaces to segment. For each B-scan, \emph{conv-s} estimates a position distribution map $P_{W \times H \times S}$ via a column-wise (A-scan) softmax function \cite{2021_He}. In particular, for each A-scan $i$, $P(Y\mid i,\,s)$ is a probability mass function (PMF) that encodes the probability of $Y$ taking A-scan position $r$ for surface $s$. 

The surface position $y^s_i$ at the A-scan $i$ of surface $s$ is then inferred as a mean value of $Y$, $y^s_i = \sum_{r}{r\cdot P(r\mid i,s)}$.
Then, anatomical layer ordering is imposed by iteratively updating the inferred layer boundary positions, as in \cite{2019_He_CONF, 2021_He}: $y^s_i = y^{s - 1}_i + \abs*{y^{s}_i - y^{s - 1}_i}_+$.

\subsubsection{Topologically correct anatomical factors} 
We then use a novel \emph{topological engine} to convert the surface probability maps $P$ to $S$ anatomical factors $M_{W\times H}$.
The \emph{topological engine} performs the column-wise top-down cumulative sum of each $P^s$, obtaining a representation $C^s$ which has values close to 0 above $y^s$ and close to 1 below it. Again, to guarantee the correct topological ordering, we iteratively update the generated surfaces: $M^s = \abs*{C^k + C^{k - 1} - 1}_+$.
Then, to have mutually exclusive spatial maps for the retinal layers, we decompose the cumulative surface maps to layer maps iteratively by setting $M^s = M^s - M^{s+1}$.
The layer maps $M$ and the texture factor are rounded to the nearest integers, forming the anatomical factors of the disentangled representation.

\subsubsection{Supervised training with layer annotations}

During training, we alternate between a supervised and a self-supervised model update, depending on the availability of layer annotations.
For supervised training, two loss terms are used: 
\begin{inparaenum}
\item $\mathcal{L}_{\mathrm{KL}}$, which describes the target distribution of  $P(Y\mid i,\,s)$, and
\item $\mathcal{L}_{\text{mse}}$, which penalizes wrongly regressed layer boundary locations with mean square error (MSE).
\end{inparaenum}
For $\mathcal{L}_{\mathrm{KL}}$ in particular, we model $P(Y\mid i,\,s)$ as a normal distribution to accommodate possible pixel-wise imprecision of the annotations. For each A-scan $i$, the distribution's mean is the reference standard position $\mu^s_{i}$ of the surface and the standard deviation $\sigma$ is a hyperparameter.
The mean Kullback-Leibler (KL) divergence\cite{1951_Kullback} is used as a loss term during the training to encourage the network to learn the target distribution \cite{2019_He_CONF}, leading to the supervised loss:

\begin{equation}
\begin{split}
   \mathcal{L}_{\text{sup}} = \lambda_1\mathcal{L}_{\mathrm{KL}} + \lambda_2\mathcal{L}_{\text{mse}} =&  \lambda_1\sum_s\sum_r\sum_{i}{P\left(r \mid i, s\right) \cdot P\left(i, s\right) \ln \frac{P\left(r\mid i, s\right)}{T\left(r\mid i,s\right)}}+\\
   &+\lambda_2\frac{1}{S\cdot W}\sum_s\sum_i\left(y^s_i - \mu^s_i\right)^2
\end{split}
\end{equation}

\noindent where $T(Y \mid i,s) \sim N(\mu^s_i \mid \sigma)$ is the target probability map containing a Gaussian normal distribution and $\lambda$ are weighting factors.

\subsubsection{Self-supervised training with anatomical priors}

We propose a set of tailored constraints to promote segmenting anatomically coherent surfaces. These are used whenever annotations are not available, and this extra supervision signal provided is especially important in scenarios with very limited training data.

\noindent 1. \textit{Topological ordering constraint}
Since retinal layers are strictly ordered, we encourage minimizing the number and magnitude of the mean topological ordering violations among A-scans: $\mathcal{L}_{\text{to}} = \frac{1}{W}\sum_s\sum_i \abs*{y^{s-1}_i - y^{s}_i}_+$.

\noindent 2. \textit{Surface continuity constraint}
To promote surface continuity, we penalize position differences between two adjacent A-scans larger than a layer specific constant $c_s$, derived from the training annotations: $\mathcal{L}_{\text{lc}} =  \frac{1}{W}\sum_s\sum_i \abs*{\abs*{y^{s}_i - y^{s}_{i+1}} - c_s}_+$.

\noindent 3. \textit{Surface slope constraint}
We encourage the curvature of the surfaces to be within an expected range, as each of them have an inherent maximum slope, assuming B-scans with low inclination. For instance, normally, the maximum slope of the BM~(Suppl. Fig.~1) is lower than of RPE in the presence of AMD. The annotated train data is used for computing a maximum expected slope $o_s$ per layer for A-scans $\delta$ pixels apart ($\delta$ is a tunable hyperparameter). Each inferred surface slope greater than $o_s$ is penalized: $\mathcal{L}_{\text{ls}} =  \frac{1}{W}\sum_s\sum_i \abs*{\frac{|y^{s}_i - y^{s}_{i+\delta}|}{\delta} - o_s}_+$.

\noindent 4. \textit{Probability distribution standard deviation constraint} 
We minimize the number of probable adjacent vertical positions for each surface on $P(Y|i, s)$ to encourage the network to settle at a specific layer boundary position and to predict a Gaussian with small variance, similarly to $\mathcal{L}_{\mathrm{KL}}$. First, we compute the standard deviation for each A-scan $i$, $\hat\sigma^s_{i}$.

We then penalize predicting locations with high uncertainty by introducing a maximum threshold for $\hat\sigma^s_{i}$: $\mathcal{L}_{\text{std}} = \sum_s\sum_i \abs*{\hat\sigma^s_{i} - t}_+$, where $t$ is a tunable hyperparameter.

\subsection{Image reconstruction}
\subsubsection{Style factor generation}
The style factor generation follows the modality factor generation in Chartsias\etal\cite{2019_Chartsias}. In particular, a Variational Autoencoder (VAE) generates the style factors, which are later used in the reconstruction part. During training, the loss term $\mathcal{L}_{z_{\mathrm{KL}}}$ \cite{2019_Chartsias} minimizes the KL divergence of the estimated Gaussian distribution from the unit Gaussian.

\subsubsection{Image decoder}
The image decoder block is similar to Chartsias\etal\cite{2019_Chartsias}, where the style and the binarized anatomical factors are combined to create an image of the same size as the input. The fusion is achieved via a model conditioned with four FiLM \cite{2018_Perez} layers. This method only scales and offsets the intensities of the anatomical factors, ensuring that no spatial information is passed through the style factors. The reconstruction loss $\mathcal{L}_{\text{rec}}$ is the Mean Absolute Error of the pixel intensities of the original and reconstructed images between the predicted ILM and BM, the top-most and the bottom-most surfaces, to avoid the influence of the noisy region outside the retina.

\par~\par
The overall cost function $\mathbf{L}$ is a composition of the style encoder loss, the reconstruction loss and the mean of the supervised $\mathcal{L}_{\text{sup}} = \lambda_1 \mathcal{L}_{\mathrm{KL}} + \lambda_2 \mathcal{L}_{\text{mse}}$ and self-supervised $\mathcal{L}_{\text{self}} = \lambda_3\mathcal{L}_{\text{to}} + \lambda_4\mathcal{L}_{\text{lc}} + \lambda_5\mathcal{L}_{\text{ls}} + \lambda_6\mathcal{L}_{\text{std}}$ loss:
    \begin{align*}
    \mathbf{L} &= \lambda_7\mathcal{L}_{z_{\mathrm{KL}}} + \lambda_8\mathcal{L}_{\text{rec}} + \frac{1}{2}\left(\mathcal{L}_{\text{sup}} + \mathcal{L}_{\text{self}}\right)
    \end{align*}

\section{Experimental results}

\paragraph{\textbf{Dataset}}

The proposed method was trained and evaluated on an internal dataset of 459 volumetric OCT scans of 459 eyes spanning intermediate and late wet-AMD real-world clinical cases. The scans were acquired with Spectralis scanners (Heidelberg Engineering) consisting of $(512 \times 19 \times 496) - (1024 \times 49 \times 496)$ voxels covering a $6 \times 6 \times 2 \text{mm}^3$ volume. For 68 volumes, 7 B-scans uniformly spaced across the volume contained manual layer annotations by experts, totaling 476 annotated B-scans. The 68 volumes were randomly split into a training set, containing 54 volumes, and into validation and test sets, each containing 7 volumes. The remaining 391 scans without annotations were used for the self-supervised training set.
\paragraph{\textbf{Training details}}

The volumes were normalized volume-wise to zero mean and unit standard deviation. During training, B-scans were randomly flipped horizontally $(p = 0.3)$ to augment the data.

The loss weight terms $\lambda_7$ and $\lambda_8$ are the same as in \cite{2019_Chartsias}. The remaining loss hyperparameters were optimized via grid-search on the validation set. In particular, these values were initially set to 1 and were either scaled up or down by 5 or 10. The network was trained for 50 epochs (approx. 1 hour of training time) and the best-performing configuration was selected, leading to $\lambda_1=\lambda_2=50, \lambda_3=\lambda_4=\lambda_5=\lambda_8=1, \lambda_6=\lambda_7=0.1$. The low value of $\lambda_6$ allows more flexibility in scenarios where a layer might be not present or invisible, and thus a larger $\hat\sigma^s_{i}$ is possible. The high values of $\lambda_1$ and $\lambda_2$ promote correct layer boundaries and allow to counter-balance the reconstruction loss, which could otherwise encourage layers with similar intensities to be in the same anatomical factor.  For $\mathcal{L}_{\text{std}}$, $\mathcal{L}_{\text{ls}}$ and $\mathcal{L}_{\mathrm{KL}}$  we used $t = 1$, $\delta = 10$ and $\sigma=0.5$, respectively. These values were also determined with grid-search after fixing all $\lambda$s.
The values of $c_s$ and $o_s$ are presented in the supplementary material.

The training was done with a batch size of 14 with 7 labeled and 7 unlabeled samples, using a MADGRAD optimizer\cite{2021_Defazio} with a learning rate of $10^{-4}$. To mitigate training instabilities, a gradient norm clipping was applied with a maximum norm of 0.5. We trained all of the methods for 300 iterations and the models with the lowest average root mean squared error (RMSE) on the validation set were picked for testing.

The training was performed in a mixed-precision setting on an Nvidia GeForce RTX 2080 Ti GPU, with an Intel Xeon Silver 4114 CPU with 16 cores, under CentOS 7, using Python 3.8.8 and PyTorch 1.8.1. The training lasted for 6 hours on this computer setup. The source code of our implementation is available at \url{http://github.com/ABotond/SD-LayerNet}. 

\paragraph{\textbf{Experiments}}

We compared our method to both supervised and a semi-supervised \emph{baselines}. The supervised baseline was the state-of-the-art boundary regression method from He\etal\cite{2021_He}, using an Attention Res-U-Net for feature extraction. To ease comparison, B-scan flattening and cropping were not performed, and instead, we use the same input as for our method. The semi-supervised baseline was SD-Net\cite{2019_Chartsias}, which classifies the pixels into surfaces classes similarly to \cite{2017_Roy}, but it is assisted with self-supervised training. We used the same optimizer and learning rate as for our method, while the other hyperparameter were kept from the original methods. 

We conducted \emph{ablation studies} on our method to assess the effect on the performance of the anatomical priors, texture factor, and image reconstruction. Further, we conducted an experiment by combining the method of He\etal{} with the additional priors.


To evaluate the methods' performance with a limited number of labeled samples, the networks were trained on a $50\%$, $25\%$, and $15\%$ randomly reduced subset of the training set, while ensuring that all original training volumes were still represented. For instance, for the $15\%$ training set a single B-scan was randomly selected from each of the annotated volumes. We conducted a statistical test with a linear mixed-effect model \cite{2014_Bates}, which takes the interdependency of the B-scans from the same volume into account.



\paragraph{\textbf{Results}}

\begin{figure}[b]
    \centering
    \includegraphics[width=0.99\textwidth]{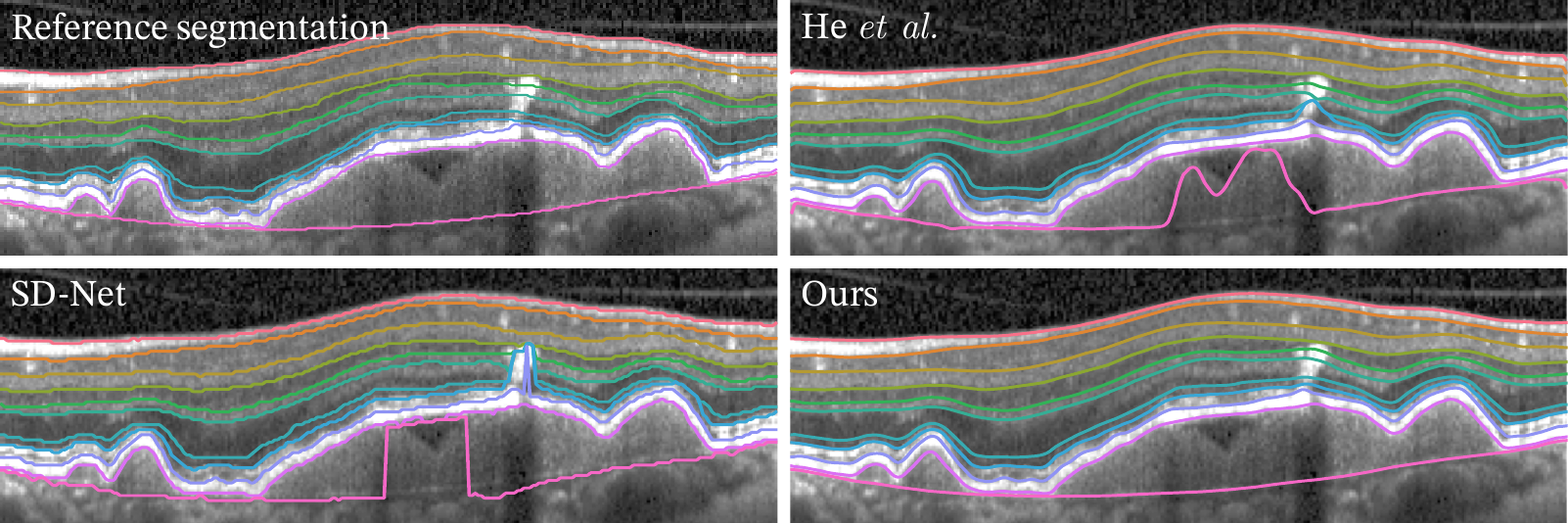}
    \caption{Manual reference segmentations and models' predictions (from upper surface ILM to bottom surface BM).}
    \label{fig:sample_segmentations}
\end{figure}

Our baseline comparison study (Table~\ref{tab:perf_comparison})  shows a significant improvement upon the two baseline methods in all subsets of the training set, with error reductions ranging from 18\% when training on the full training set up to a 30\% lower error rate when training on a reduced training set. In particular, our method, when trained on 15\% of the training set, performed only 7\% worse than the He\etal{} method trained on the full set, while yielding more consistent results, as indicated by the lower standard deviation. Also, the introduced prior knowledge allows for a more robust segmentation, specially on pathological regions (Figure~\ref{fig:sample_segmentations}). These results support our hypothesis that incorporating a self-supervised paradigm, with disentangled learning, can considerably improve the data efficiency of layer regression methods.

\begin{figure}[tb]
\includegraphics[width=0.98\textwidth]{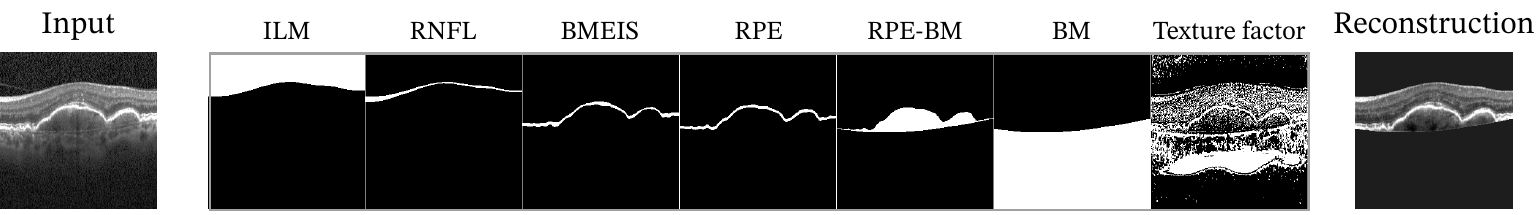}
\caption{A sample subset of the anatomical factors (six out of 12), and the texture factor for an intermediate AMD case.} 
\label{fig:t-texture-factor}
\end{figure}

The ablation study (Table~\ref{tab:perf_comparison})  shows that the newly proposed anatomical priors $\mathcal{L}_{\text{self}}$ and the texture factor contributed to lower the segmentation error. Indeed, $\mathcal{L}_{\text{self}}$ improves the performance and consistency both of our method and of He\etal, indicating that this additional information is guiding the network to a better local optimum. Also, the texture factor (e.g. Figure~\ref{fig:t-texture-factor}), alleviates the reconstruction burden and improves segmentation by enabling the anatomical factors to map solely the expected position and overall intensity of the layers.

\begin{table}[t]
    \centering
    \caption{Average B-scan-wise root mean square error ($\upmu$m) and standard deviation of the proposed method and the baselines for different percentages of the training set. *~indicates statistical difference to our method ($p<0.05$)}
\begin{tabular}{|l|c|c|c|c|}
\hline
\diagbox[width=3.2cm]{\textbf{Method}}{\textbf{Subset size}} & \textbf{15\%}            & \textbf{25\%}            & \textbf{50\%}            & \textbf{100\%}           \\ \hline
\textbf{SD-Net} $\approx$ \cite{2019_Chartsias}                                              & $14.10 \pm 11.26$*         & $11.76 \pm 8.06$*         & $9.77 \pm 5.43$*          & $9.44 \pm 5.20$*           \\ \hline
\textbf{He\etal} $\approx$ \cite{2021_He}                                              & $10.48 \pm 5.95$*         & $10.16 \pm 6.96$*         & $7.6 \pm 4.18$*           & $7.06 \pm 3.44$*          \\ \hline
\textbf{He\etal + $\mathcal{L}_{\text{self}}$}                                             & $10.35 \pm 2.49$*         & $8.63 \pm 1.72$*         & $7.38 \pm 2.03$*           & $6.86 \pm 1.24$*          \\ \hline
\textbf{Ours without $conv$-$t$}                                                & $10.28 \pm 4.34$*  & $8.64 \pm 2.33$*  & $7.41 \pm 2.03$*  & $7.11 \pm 1.87$*           \\ \hline
\textbf{Ours without $\mathcal{L}_{\text{self}}$}                         & $8.12 \pm 4.44$          & $7.04 \pm 3.57$          & $6.55 \pm 2.47$          & $\mathbf{5.79 \pm 1.33}$ \\ \hline
\textbf{Ours}                                                & $\mathbf{7.57 \pm 2.94}$ & $\mathbf{6.95 \pm 1.69}$ & $\mathbf{6.38 \pm 1.53}$ & $5.82 \pm 1.32$          \\ \hline
\end{tabular}
    
    \label{tab:perf_comparison}
\end{table}

\section{Conclusion}

Our data-efficient layer segmentation model outperforms the current state of the art by a large margin on our test set. 
In particular, the proposed fully-differentiable topological engine allows to use both 1D surface and 2D {layer} information in a coupled fashion to train the model. 
This module and the anatomical priors encoded on dedicated loss terms allow to efficiently make use of large amounts of unlabeled data by guiding the network to create mutually exclusive anatomically constrained spatial representations of the layers to reconstruct the input image.
The applicability of the proposed method is not restricted to retinal layer segmentation but can be applied to any nested anatomy and where the thickness of a tissue is measured (e.g. inner and outer vessel lumen wall, cardiac wall, knee cartilage, etc.).

Future research aims at exploiting the style and texture factors to improve the generalization of the model. In particular, forcing the texture factor to have complete anatomical independence and a more dedicated use of the style factor may allow this method to be used for other tasks such as denoising, synthetic B-scan generation for data augmentation, and improved performance (or even image conversion) across different imaging devices.

\section*{Acknowledgements}
The financial support by the the Christian Doppler Research Association, Austrian Federal Ministry for Digital and Economic Affairs, the National Foundation for Research, Technology and Development, and Heidelberg Engineering is gratefully acknowledged.

%
%
%
\bibliographystyle{splncs04}
\bibliography{paper1998}

\end{document}